# Spin Squeezing Induced Enhancement of Sensitivity of an Atomic Clock using Coherent Population Trapping

Jinyang Li[1], Gregório R. M. da Silva[1], Schuyler Kain[1], Gour Pati[2], Renu Tripathi[2], Selim M. Shahriar[1,3]

[1] Department of Physics and Astronomy, Northwestern University, Evanston, IL 60208, USA
[2] Division of Physics, Engineering, Mathematics and Computer Science, Delaware State University, Dover, DE 19901, USA
[3] Department of Electrical and Computer Engineering, Northwestern University, Evanston, IL 60208, USA

## Abstract

The coherent population trapping (CPT) effect is used for making compact atomic clocks. There are two types of CPT clocks: the one in which the Raman beams are applied continuously and the one in which two CPT pulses separated by a dark period are applied (Ramsey scheme). It is obvious that the technique of spin squeezing can only be applied to the Ramsey CPT clock to enhance the sensitivity. However, it is not apparent how to adapt to the CPT clock the protocols for the microwave clock using one-axis-twist squeezing (OATS), since the Ramsey CPT clock is not trivially equivalent to the Ramsey microwave clock. In this paper, we show explicitly how to adapt two protocols using OATS, namely the Schrödinger cat state protocol (SCSP) and the generalization thereof, and the echo squeezing protocol (ESP), to the CPT clock. The ESP magnifies the phase shift by a factor of $\sqrt{N/\mathrm{e}}$, while the SCSP magnifies the phase shift by a factor of $N/2$, making it able to achieve a higher sensitivity in the presence of excess noise.

## 1. Introduction

In an alkali atom-based clock employing the process of coherent population trapping (CPT), two laser beams are used to excite a resonant Raman transition that couples the two hyperfine ground states [1,2,3,4,5,6,7]. Fundamentally, it is similar to a microwave atomic clock. However, the use of lasers to excite the microwave transition makes it unnecessary to use microwave fields directly. As a result, it is possible to realize a CPT clock with very small form factors [5,6]. For example, the only chip-scale atomic clock available commercially is a CPT clock [8]. As such, there is a significant interest and on-going effort in further development of the CPT clock, using Rb or Cs [9,10,11,12,13,14,15,16,17,18,19,20,21,22,23]. In particular, efforts are currently underway to make compact CPT clocks using cold atoms released from a magneto-optical trap [14,24,25]. The sensitivity of a CPT-based clock is expected to be similar to that of a microwave clock, for similar number of atoms for interrogation. There is some concern that fluctuations in light shifts in a CPT clock may limit its sensitivity; however, significant work has been carried out to suppress this effect strongly [14,26,27,28]. The current CPT clocks are generally not limited by the quantum projection noise. For example, the cold atom-based CPT clock reported in Ref. [14], which makes use of 5 million atoms, reports that the sensitivity achieved at one second averaging time is lower than the standard limit of the quantum projection noise by a factor of 50. Conventional techniques of spin squeezing that only suppresses the quantum projection noise is thereby unlikely to improve the sensitivity of a such a clock. However, recently developed protocols [ 29 , 30 , 31 ], based on one-axis-twist squeezing (OATS) [32,33,34,35,36,37,38], makes use of phase magnification rather than suppression of quantum noise for enhancing sensitivity. When such a protocol is employed, it is possible to increase the sensitivity of a sensor even when the excess noise is significantly larger than the standard limit of the quantum projection noise. As such, there is an interest in exploring the use of this type of spin-squeezing protocols for enhancing the sensitivity of CPT based atomic clocks.

To investigate the feasibility of applying spin squeezing to a CPT clock, it is necessary to distinguish between two types of CPT clocks, namely the type in which the Raman beams are applied continuously (single zone scheme) and the type in which two CPT pulses separated by a dark period are applied (Ramsey scheme). The CPT process itself makes use of spontaneous emission, which is a highly incoherent process, while the process of ideal spin squeezing makes use of a fully coherent non-linear atom-field interaction. For the CPT clock employing the single zone scheme, spontaneously emission is continuously present if the clock frequency is detuned, leaving no time for the coherent process of spin squeezing. On the contrary, in the Ramsey CPT clock, there is a dark period between the two CPT pulses during which the atoms are coherent. Consequently, spin squeezing can be applied to the CPT clock to enhance the sensitivity. Nevertheless, the adaptation necessary for the CPT clock is not apparent because the Ramsey CPT clock is not trivially equivalent to the Ramsey conventional microwave clock. It is well known [39] that there is an important difference between the dark state produced at the end of the saturating CPT pulse (which is the first of the two CPT pulses) and the state produced with a microwave $\pi/2$ pulse. To illustrate this difference, we adopt the notation that all atoms are represented as pseudo-spins, with the $\pm z$-directions being the two hyperfine ground states used for the clock transition. Under the microwave excitation with all the pseudo-spins starting in the $z$-direction, the $\pi/2$ pulse produces a state which is aligned in the $y$-direction (in the rotating wave frame). However, after a saturating CPT pulse that produces the dark state, the pseudo-spins are aligned in the $x$-direction (in the same rotating wave frame).

This difference has a significant consequence for adapting the protocols using one-axis-twist squeezing (OATS) to the CPT clock, because the pairing of the orientation of the pseudo-spins and the rotation axis of all the pulses will considerably affect the degree of achievable enhancement in sensitivity. One example is the choice of the rotation axis of the auxiliary and the inverse auxiliary pulses. The explicit OATS protocols we have considered in this manuscript, namely the Schrödinger cat state protocol (SCSP) [29], the generalization thereof [31], and the echo squeezing protocol (ESP) [30], employ auxiliary rotations and inversions of the rotations. As we have also shown in Ref. [29] and [31], different axes must be used for these auxiliary rotations and inverse rotations for different parities of the number of atoms for achieving the Heisenberg limit using the SCSP. For the ESP [30], non-zero signal will be observed only for one choice of the rotation axis of the auxiliary and the inverse auxiliary rotations. Zero signal will be obtained if the wrong rotation axis is chosen.

Given these facts, it is not *a priori* clear how the protocols of interest (namely the SCSP, the generalization thereof, and the ESP) need to be modified so that they can produce optimal enhancement in sensitivity for the CPT clock. Answering this question quantitatively is the main contribution of and the primary motivation for this paper.

The rest of the paper is organized as follows. In Section 2, we review the conventional Ramsey CPT clock. In Section 3, we propose a more efficient optical pumping scheme for the first CPT pulse. In Section 4, we show the application of the Schrödinger cat state protocol and the generalization thereof to the CPT clock. In Section 5, we show the application of the echo squeezing protocol to the CPT clock, followed by the conclusion and discussion in Section 6.

## 2. Conventional CPT clock

Before we discuss the conventional CPT atomic clock, we first review a two-level system in order to establish our language and notation for the remainder of the manuscript. A two-level system is equivalent to a spin-1/2 spinor, with the spin operator $\mathbf{s} \equiv (s_x, s_y, s_z)$. The two eigenstates of $s_z$ are denoted as $|\uparrow\rangle$ and $|\downarrow\rangle$ with eigenvalues of $1/2$ and $(-1/2)$. The state of a two-level system can be described by a point on the Bloch sphere. A point on the Bloch sphere can be characterized by two angle parameters: $\theta$ and $\phi$. The $j$-th spinor in the state corresponding to such a point on the Bloch sphere is defined as [32] $\left|(\theta,\phi)_j\right\rangle \equiv \cos(\theta/2)|\uparrow_j\rangle + e^{i\phi}\sin(\theta/2)|\downarrow_j\rangle$. A coherent spin state (CSS) characterized by the parameters $\theta$ and $\phi$ is defined as a state of $N$ atoms with each atom in the state $\left|(\theta,\phi)_j\right\rangle$, that is

$$|\theta,\phi\rangle \equiv \bigotimes_{j=1}^{N} \left|(\theta,\phi)_j\right\rangle.$$

In the conventional CPT clock, we use nominally three-level atoms and a pair of Raman beams. The three states consist of two ground states, denoted as $|\uparrow\rangle$ and $|\downarrow\rangle$, and an excited state, denoted as $|e\rangle$. The $|\uparrow\rangle$ and $|\downarrow\rangle$ are the $m_F = 0$ Zeeman sublevels in the two hyperfine states of an alkali atom. Each Raman beam couples one of the ground states to the excited state, which can decay to the two ground states, as well as the other Zeeman sublevels. There are potential complications due to the presence of the $m_F \neq 0$ sublevels in the ground states and multiple Raman transition channels. We will discuss these issues in Section 3.

Under the electric-dipole approximation and the rotating-wave approximation, the Raman interaction in the rotating-wave basis can be described by a Hamiltonian in the basis of $\{|\uparrow\rangle, |e\rangle, |\downarrow\rangle\}$ ($\hbar = 1$), as follows

$$H = \frac{1}{2}\begin{bmatrix} \delta & \Omega_\uparrow & 0 \\ \Omega_\uparrow & -2\Delta & \Omega_\downarrow e^{-i\phi_0} \\ 0 & \Omega_\downarrow e^{i\phi_0} & -\delta \end{bmatrix} \quad (1)$$

Here, $\delta \equiv (\delta_\uparrow - \delta_\downarrow)$ is the difference detuning, and $\Delta \equiv (\delta_\uparrow + \delta_\downarrow)/2$ is the common detuning, where $\delta_\uparrow (\delta_\downarrow)$ is the detuning for the laser field coupling the $|\uparrow\rangle (|\downarrow\rangle)$ to the excited state, $\Omega_{\uparrow(\downarrow)}$ is the Rabi frequency of the Raman beam coupling $|\uparrow\rangle (|\downarrow\rangle)$ to the excited state, and $\phi_0$ is the initial phase difference between Raman beams. The scheme is illustrated in Figure 1 (a).

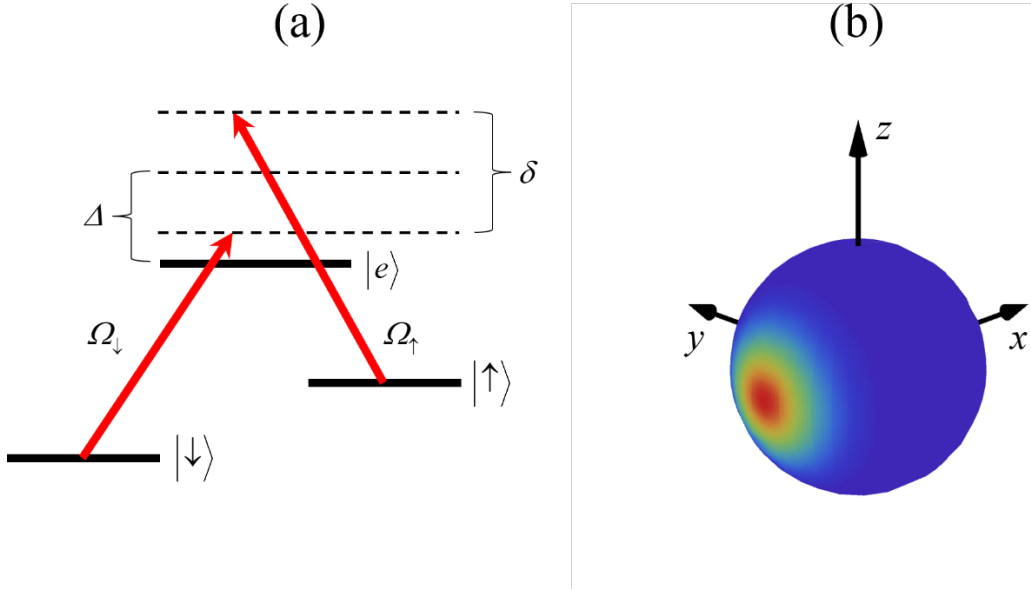

Figure 1: (a) Scheme of the Raman beams for CPT. We use nominally three-level atoms and a pair of Raman beams. The three states consist of two ground states, denoted as $|\uparrow\rangle$ and $|\downarrow\rangle$, and an excited state, denoted as $|e\rangle$. The $|\uparrow\rangle$ and $|\downarrow\rangle$ are the $m_F = 0$ Zeeman sublevels in the two hyperfine ground states of an alkali atom. Each Raman beam couples one of the ground states to the excited state, which can decay to the two ground states, as well as the other Zeeman sublevels. Here, $\delta \equiv (\delta_\uparrow - \delta_\downarrow)$ is the difference detuning, and $\Delta \equiv (\delta_\uparrow + \delta_\downarrow)/2$ is the common detuning, where $\delta_\uparrow (\delta_\downarrow)$ is the detuning for the laser field coupling the $|\uparrow\rangle (|\downarrow\rangle)$ to the excited state, $\Omega_{\uparrow(\downarrow)}$ is the Rabi frequency of the Raman beam coupling $|\uparrow\rangle (|\downarrow\rangle)$ to the excited state (b) Husimi quasi-probability distribution for the CPT dark state of the Raman beams for $\phi_0 = 0$.

If $\delta = 0$, this Hamiltonian has an eigenstate of the form $\left(\Omega_\downarrow, 0, -e^{i\phi_0}\Omega_\uparrow\right)$, which is a superposition of only $|\uparrow\rangle$ and $|\downarrow\rangle$. Because this state is not coupled to the excited state, it is called a dark state. We can see that the relative phase between $|\uparrow\rangle$ and $|\downarrow\rangle$ in this dark state is $\pi + \phi_0$, where $\phi_0$ is the initial (time-independent) relative phase between the Raman beams. We assume without loss of generality that $\phi_0 = 0$. For simplicity of analysis, we further assume that $\Omega_\uparrow = \Omega_\downarrow$. In this case, the dark state expressed with the Bloch sphere coordinates is $\left|\frac{\pi}{2}, \pi\right\rangle$, which is an eigenstate of $s_x$, with an eigenvalue of $(-1/2)$. This is shown in Figure 1(b), where we have plotted the Husimi Quasi-Probability Distribution (QPD) for the spinors. The orthogonal state, called the bright state, has coordinates $\left|\frac{\pi}{2}, 0\right\rangle$, which is also an eigenstate of $s_x$ with an eigenvalue of $1/2$.

In a typical Ramsey CPT clock, the atoms are subjected to two pulses separated in time. The duration of the first one, called the saturating pulse, is chosen to be long enough to ensure that the system is optically pumped into the dark state completely. The time scale for pumping all the atoms to the CPT dark state is about $10\left(\Omega^2/2\pi\Gamma\right)^{-1} = 1.6$ µs for $\Omega = \Gamma$, where $\Omega$ is the Rabi frequency of the transition from the bright state to the excited state, and $\Gamma$ is the decay rate of the excited state. This occurs independent of the initial state of the atoms. The operation of the Ramsey CPT clock is restricted to values of $\delta$ much smaller than the rate of optical pumping into the dark state, so that the assumption of the system being fully in the dark state at the end of the first pulse remains valid.

It should be noted that the dark state resulting from the first pulse in the Ramsey CPT clock can also be generated using a microwave field only, if all atoms are initially in, for example, the $|\uparrow\rangle$ state. To generate the state $\left|\frac{\pi}{2},\pi\right\rangle$ with the microwave field resonant with the energy difference between $|\uparrow\rangle$ and $|\downarrow\rangle$, we would need to apply a π/2-pulse corresponding to the Hamiltonian

$$H = \frac{1}{2}\begin{bmatrix} 0 & e^{i\phi_m}\Omega_0 \\ e^{-i\phi_m}\Omega_0 & 0 \end{bmatrix} = -\Omega_0 S_y; \quad \phi_m = 90° \tag{2}$$

where $\phi_m = 90°$ is the initial (time-independent) phase of the microwave field. Physically, this means that the initial phase of the microwave field must differ from the initial phase difference between the two Raman beams by 90°, as demonstrated experimentally by us previously [39]. An important consequence of this fact is that phases of the pulses in various versions of the OATS protocols designed for the microwave clock must be modified to adapt to the CPT clock. As mentioned earlier, this modification is not trivial, due to the presence of auxiliary rotations and inversions thereof in the versions of the OATS protocols of interest. We will specify shortly the modifications necessary for these protocols.

For the Ramsey CPT Clock, the first saturating pulse is followed by a dark period. With $\Omega_\uparrow = \Omega_\downarrow = 0$, the Hamiltonian for the dark zone in the basis of $\{|\uparrow\rangle, |\downarrow\rangle\}$ is then $s_z\delta$, and the corresponding propagator is $\exp(-is_z\delta t)$. Thus, during the dark period the state will rotate around the z-axis of the Bloch sphere if the difference detuning $\delta$ is non-zero. The final state after the dark period, for duration $T$, is $\left|\frac{\pi}{2},\pi+T\delta\right\rangle$. The second pulse in the Ramsey CPT clock serves as a probe to detect the population of the bright state $\left|\frac{\pi}{2},0\right\rangle$ [40]. This is equivalent to measuring the operator $S_x$, defined as the sum of the single atom spin operator $s_x$ of all the atoms. Therefore, no additional probe beam is needed, contrary to the conventional microwave Ramsey atomic clock. The signal is proportional to $\langle S_x \rangle = -(N/2)\cos(T\delta)$, where $N$ is the number of atoms. The standard deviation of $S_x$, representing the quantum projection noise (QPN) [41], is $\Delta S_x \equiv \sqrt{\langle S_x^2 \rangle - \langle S_x \rangle^2} = (\sqrt{N}/2)|\sin T\delta|$. As such, the QPN-limited uncertainty in the measurement of the clock detuning can be expressed as $\Delta \delta = |\partial \delta / \partial \langle S_x \rangle| \Delta S_x = (\sqrt{N}T)^{-1}$. It should be noted that the corresponding QPN-limited uncertainty for the detuning of a microwave-based Ramsey clock has the same expression. The stability represented by this expression is the so-called standard quantum limit (SQL).

It is well known that by making use of entangled states it is possible to exceed the SQL, and the upper bound of sensitivity is the Heisenberg Limit, which represents an improvement in clock stability by a factor of $\sqrt{N}$. For a clock employing a hundred million atoms, for example, this would represent an enhancement in sensitivity by a factor of ten thousand. In recent years, significant investigations, theoretically as well as experimentally, have been carried out to explore the feasibility of enhancing the sensitivity of microwave-based Ramsey clocks beyond the SQL, using the technique of spin-squeezing [32]. To date, the best result achieved is a suppression of signal variance by a factor of ~100, which represents an improvement in sensitivity by a factor of 10 [42]. While this is far below the Heisenberg limit, efforts are continuing to identify a realistic technique that would lead to a much larger degree of enhancement in sensitivity. In particular, we have recently shown that using the OATS employing an optical cavity, it may be possible to achieve a sensitivity close to the Heisenberg limit, even in the presence of significant excess noise due to cavity decay and residual spontaneous emission [29]. In Ref. [29], we considered the application of this technique to an atomic interferometer and a microwave-based Ramsey clock. We now show that, indeed, this approach can also be applied to the Ramsey CPT clock, with proper adaptations.

## 3. Optical pumping scheme for maximizing the CPT clock signal

Before presenting the details of how to apply spin-squeezing to the Ramsey CPT clock, we address the issues pertaining to the presence of $m_F \neq 0$ Zeeman substates and multiple Raman transition channels, as alluded to earlier. In the case of a conventional Ramsey CPT clock, these issues simply affect the overall fringe contrast in most case, and therefore not necessarily critical. However, when considering the use of spin-squeezing, the existence of the additional Zeeman sublevels and Raman transitions must be addressed carefully, since the spin-squeezing process must address the quantum state of all the atoms.

During the saturating pulse, some atoms will be pumped to the $m_F \neq 0$ Zeeman substates in both hyperfine levels. Conventionally, one uses a bias magnetic field to prevent the $m_F \neq 0$ sublevels in the ground states from contributing to the final signal, thus sacrificing the atoms lost to these sublevels. Here, we propose a scheme of applying two optical pumping beams throughout the saturating pulse to prevent such a loss. To illustrate this concept, we consider as an example the $^{87}$Rb system shown in Figure 2(a). The two optical pumping beams are both π-polarized. The first one (red arrows) is resonant with the transition from $F=1$ to $F'=1$ in the $D_1$ manifold, and the second one (blue arrows) is resonant with the transition from $F=2$ to

$F' = 2$, also in the D₁ manifold. Noting that the π-transition is forbidden for any $m_F = 0$ Zeeman substates if $\Delta F = 0$, we can see that in the presence of these two optical pumping beams only, the $m_F = 0$ Zeeman substates of both $F = 1$ and $F = 2$ hyperfine ground states are decoupled from any excited Zeeman sublevel. Thus, all atoms would end up in these Zeeman sublevels due to spontaneous emission, in the limit where we can ignore collisional decay from these states. While there are other optical pumping schemes that would achieve the same goal for some particular alkali atom, the approach described here would work for any alkali atom.

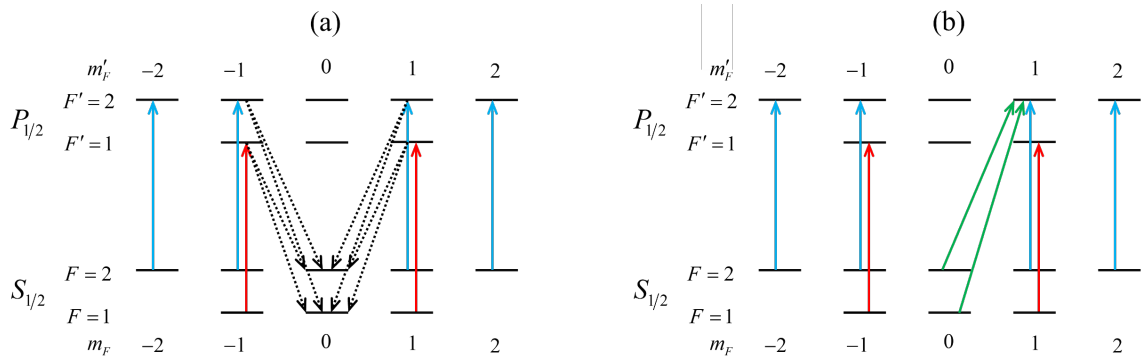

Figure 2: Optical pumping scheme for $^{87}$Rb. Two π-polarized optical pumping beams are applied. The red arrows show the optical pumping beam resonant with the transition from $F = 1$ to $F' = 1$ in the D₁ line, and the blue arrows show the beam resonant with the transition from $F = 2$ to $F' = 2$ in the D₁ line. In (a) the dashed black arrows show the decay channel to the $m_F = 0$ Zeeman substates in the ground state. Other decay channels are not shown here. In (b), no decay channel is shown. The green arrows are the CPT Raman beams.

Consider next the presence of the two laser beams to be used for generating the CPT dark state, as shown in Figure 2(b). Specifically, we consider the case where these two beams are each right circularly polarized, and tuned to the $F = 1$ to $F' = 2$ and the $F = 2$ to $F' = 2$ transitions in the D₁ manifold. The three-level system involving the two $m_F = 0$ ground Zeeman sublevels now involve the $m'_F = 1$ Zeeman sublevel of the $F' = 2$ hyperfine state as the excited state. As such, the CPT dark state, being a superposition of the two $m_F = 0$ ground Zeeman sublevels in the ground state, would be a stable dark state containing all the atoms at the end of the saturating pulse.

We should note that in this discussion we have ignored the off-resonant three-level system resulting from the coupling of the two $m_F = 0$ ground Zeeman sublevels to the $m'_F = 1$ Zeeman sublevel of the $F' = 1$ hyperfine state. However, it is easy to see that the CPT dark state corresponding to the case where $F' = 2, m'_F = 1$ is the excited state is identical to the one corresponding the case where $F' = 1, m'_F = 1$ is the excited state, since the ratios of the dipole matrix elements (including signs) for the two legs of the Raman transition are identical in these two cases [43]. Furthermore, we note that the frequency separation (815 MHz) between the $F' = 1$ and $F' = 2$ state is very large compared to typical Rabi frequencies, and use of cold atoms with a very small Doppler shift spread is envisioned for the spin-squeezed CPT clock. Therefore, in modeling the behavior of this system, it is adequate to consider only the three-level system in which the excited state is the $F' = 2, m'_F = 1$ Zeeman sublevel. We assume that the beam constituting one leg of the Raman excitation is off-set phase locked to the beam on the other leg, by using a reference voltage-controlled oscillator (VCO). In what follows, we will refer to this as the clock VCO. Nominally, its frequency is tuned to the energy difference between the $F = 2, m_F = 0$ state and the $F = 1, m_F = 0$ state. During the operation of the CPT clock, the frequency of this VCO will be stabilized using a feedback signal, and the output of this VCO will represent the CPT clock. Furthermore, we note that the generation of the CPT dark state would occur independent of the initial populations of the Zeeman sublevels and the coherence condition of the atoms in the $5S_{1/2}$ state [40].

## 4. Schrödinger cat CPT clock

There are many different definitions of a Schrödinger cat state in the context of entangled optical fields and in the arena of spin squeezed atomic states. Here, we define the Schrödinger cat (SC) state as one that is a maximally entangled state represented by a superposition of two orthogonal coherent spin states (CSS). The orientation of the SC state depends on the nature of the two CSS's [29,44]. For example, a $z$-directed SC state is a superposition of a state in which all atoms are in the $|\uparrow\rangle$ state, and another state in which all atoms are in the $|\downarrow\rangle$ state. It has been shown [29,44,45,46,47,48,49] that such an SC state can be generated with OATS. However, the orientation of the SC state depends on the parity of $N$ [29,44]. As such, any protocol involving the SC state generated in this manner must have a specified parity. For experiments involving a few ions, for example, this is not a problem [47,49]. However, it can be a real challenge for a clock employing a larger number of neutral atoms. Consider, for example, a scheme wherein the atoms are first caught in a MOT and then released for interrogation. In this case, the parity of $N$ is expected to be random, with equal probability of being even or odd. In a pair of recent papers [29,44], we have shown how to overcome this problem. Specifically, we have shown that if a protocol is designed for a particular parity, the derivative of the signal for the other parity with respect to the difference detuning is either zero or negligibly small (depending on which version of the detection scheme is used). As a result, the net effect of the randomness of the parity is a reduction in the number of atoms by a factor of 2. Thus, we show that under ideal condition, the sensitivity is enhanced by a factor of $\sqrt{N/2}$, which is the Heisenberg limit within a factor of $\sqrt{2}$. In Refs. [29] and [44], we considered two different systems: a light pulse atomic interferometer and a microwave clock. We now show how to modify this scheme for the CPT clock.

The steps in the protocol for the SC-CPT clock are illustrated schematically in Figure 3. The pulse sequence is shown in Figure 3 (a), and the Husimi QPDs after each step are shown in Figure 3 (b). For the Schrödinger cat CPT clock (SC-CPT clock), the first step is the preparation of the atoms in the CPT dark state, using a saturating pulse of the Raman beams. As noted above, this dark state can be expressed as the coherent spin state denoted as $\left|\frac{\pi}{2}, \pi\right\rangle$.

Here, for concreteness, we have assumed that prior to the start of the saturating CPT pulse (shown in the first block in Figure 3(a)), the atoms are all in the $|\uparrow\rangle$ state, as shown in Figure 3(b1). However, as discussed earlier, the system would evolve into the CPT dark state, as shown in Figure 3 (b2), regardless of the initial state. After the atoms are prepared in the CPT dark state, we squeeze them with the one-axis-twist Hamiltonian, $H = \chi S_z^2$, as indicated in the second block in Figure 3(a), for a period of time, $t$, such that $\mu \equiv \chi t = \pi/2$, to generate a Schrödinger cat (SC) state, which is shown in Figure 3(b3). One-axis-twist squeezing can be realized using a cavity, as mentioned in the introduction. We have shown an estimation of the squeezing time needed for producing the Schrödinger cat state in Appendix A of Ref. [29]. In that example, $(\delta_c/2\kappa)$ is set at 100 (where $\delta_c$ is the detuning of the probe beam to the cavity resonant frequency and $\kappa$ is the cavity decay rate), the mode area is chosen to be $(200\ \mu\text{m})^2$, and the cavity mirror transmittivity to be $10^{-4}$, which corresponds to a single atom cooperativity of 0.9. Under these conditions, an input power of 10 mW will give a value of the parameter in the Hamiltonian of $\chi = 100\ \mu\text{s}^{-1}$. In this case, the interaction time needed for producing the

Schrödinger state, which requires $\mu = \pi/2$, is 0.15 μs. This state has the form $\frac{1}{\sqrt{2}}\left[\left|\frac{\pi}{2},0\right\rangle - (-1)^{N/2} i \left|\frac{\pi}{2},\pi\right\rangle\right]$ for even $N$, and $\frac{1}{\sqrt{2}}\left[\left|\frac{\pi}{2},\frac{\pi}{2}\right\rangle - (-1)^{N/2} i \left|\frac{\pi}{2},\frac{3\pi}{2}\right\rangle\right]$ for odd $N$. In the experimental situation envisioned here, atoms will be released from a trap before applying the first CPT pulse. As such, $N$ will be even or odd with equal probability when the experiment is repeated many times. We will address the issue of what the net enhancement in sensitivity would be when averaging over both the even-$N$ and odd-$N$ trials later. In what follows, we first constrain our discussion of the protocol that is optimized for odd $N$, to be followed by a summary of what would happen under this protocol for the case of even $N$. Of course, it is also possible to construct a protocol that is optimized for even $N$, and determine the behavior of that protocol for the case of odd $N$. The results of these two protocols are essentially symmetric.

Upon completion of the one-axis-twist squeezing process with $\mu = \pi/2$, the state (for odd $N$) is maximally entangled, and can be thought of as a Schrödinger cat state in the *y*-direction, as indicated in Figure 3 (b3). However, in order to generate the desired $N$-fold phase amplification, it is necessary to have a Schrödinger cat state in the *z*-direction [29,44]. This is easily done by applying an auxiliary rotation. Specifically, we apply a pulse, as shown in the third block of Figure 3(a), that rotates the quantum state around the *x*-axis by an angle of π/2. This π/2-pulse can be realized using a resonant microwave field, or, equivalently, a pair of two-photon resonant Raman beams that are highly detuned optically [40]; the choice of either approach would be dictated by experimental constraints. For concreteness of discussion, we will assume that a microwave field would be used for this purpose. This microwave field will have to have the same phase as that of the clock VCO [50]. The resulting state, shown in Figure 3 (b4), is expressed as $\frac{1}{\sqrt{2}}\left[|\Uparrow\rangle - (-1)^{(N-1)/2} i |\Downarrow\rangle\right]$ where $|\Uparrow(\Downarrow)\rangle \equiv \bigotimes_{j=1}^{N}|\uparrow(\downarrow)\rangle$. This is followed by a dark period of duration $T$, which can be represented by the propagator $\exp(-i\delta T S_z)$. At the end of the dark period, the state, shown in Figure 3(b5), becomes $\frac{1}{\sqrt{2}}\left[|\Uparrow\rangle - (-1)^{(N-1)/2} i e^{iN\delta T} |\Downarrow\rangle\right]$.

Before further processing, it is necessary to reverse the action of the auxiliary rotation. This is accomplished by applying another microwave $\pi/2$-pulse, indicated by the fourth block in Figure 3(a), that is out of phase with respect to the field used for the auxiliary rotation (or, equivalently, a $3\pi/2$-pulse with a field that is in phase with the field used for the auxiliary rotation). The propagator for this anti-auxiliary-rotation pulse can be expressed as $\exp(i(\pi/2)S_x)$. The resulting state is shown in Figure 3(b6). Finally, it is also necessary to undo the effect of the squeezing pulse. This is accomplished by applying an un-squeezing pulse corresponding to $\mu = (-\pi/2)$ [51]. Experimentally, the reversal of the sign is produced by changing the sign of the detuning of the OATS probe beam with respect to the cavity resonance. The resulting state, shown in Figure 3(b7), can be expressed as $\frac{1}{\sqrt{2}}\left[\sin\frac{N\delta T}{2}\left|\frac{\pi}{2},0\right\rangle + (-1)^{(N-1)/2}\cos\frac{N\delta T}{2}\left|\frac{\pi}{2},\pi\right\rangle\right]$.

Once this is completed, we apply the second CPT pulse, indicated by the last block in Figure 3(a), to measure the operator $S_x$ [52]. Noting the relations $S_x\left|\frac{\pi}{2},0\right\rangle = \frac{N}{2}\left|\frac{\pi}{2},0\right\rangle$, $S_x\left|\frac{\pi}{2},\pi\right\rangle = -\frac{N}{2}\left|\frac{\pi}{2},\pi\right\rangle$, and $\left\langle\frac{\pi}{2},\pi\left|\frac{\pi}{2},0\right\rangle\right. = 0$, it is easy to show that

$$\langle S_x \rangle = \frac{N}{2}\sin^2\frac{N\delta T}{2}\left\langle\frac{\pi}{2},0\left|\frac{\pi}{2},0\right\rangle\right. - \frac{N}{2}\cos^2\frac{N\delta T}{2}\left\langle\frac{\pi}{2},\pi\left|\frac{\pi}{2},\pi\right\rangle\right. = -\frac{N}{2}\cos N\delta T \qquad (3)$$

and

$$\Delta S_x \equiv \sqrt{\langle S_x^2\rangle - \langle S_x\rangle^2}$$

$$= \sqrt{\frac{N^2}{4}\sin^2\frac{N\delta T}{2}\left\langle\frac{\pi}{2},0\bigg|\frac{\pi}{2},0\right\rangle + \frac{N^2}{4}\cos^2\frac{N\delta T}{2}\left\langle\frac{\pi}{2},\pi\bigg|\frac{\pi}{2},\pi\right\rangle - \left(\frac{N}{2}\cos N\delta T\right)^2} \quad (4)$$

$$= \sqrt{\frac{N^2}{4}(1-\cos^2 N\delta T)} = \frac{N}{2}|\sin N\delta T|$$

In the absence of the additional pulses used for squeezing, auxiliary rotation, anti-auxiliary-rotation and unsqueezing, that is, for a conventional Ramsey CPT clock, it is well known that the resulting signal can be expressed as $\langle S_x\rangle = (-N/2)\cos(\delta T)$, and that the noise can be expressed as $\Delta S_x = (\sqrt{N}/2)|\sin \delta T|$, assuming perfect quantum efficiency of detection. For the Schrödinger cat state protocol (SCSP), the phase shift caused by the clock detuning would be magnified by a factor of $N$, and the noise at zero detuning would be magnified by a factor of $\sqrt{N}$ [29]. In the absence of excess noise, the uncertainty of the measurement is $\Delta\delta = \Delta S_x / |\partial_\delta\langle S_x\rangle| = (NT)^{-1}$, which reaches the Heisenberg limit. As shown in Ref. [31], the phase magnification of a protocol indicates its ability to enhance the sensitivity in the presence of excess noise comparable or even larger than the standard limit of the quantum projection noise. Due to a phase magnification by a factor of $N$, the SCSP for odd $N$ can enhance the sensitivity substantially even in the presence of significant excess noise.

If the value of $N$ is even (while still using the protocol optimized for odd $N$), in the proximity of zero difference detuning, the signal can be approximated as [29,44] $\langle S_x\rangle = -(N/2)\cos\sqrt{N}\delta T$, and the standard deviation of $S_x$ can be approximated as $\Delta S_x = (N/2)|\sin\sqrt{N}\delta T|$. Therefore, the uncertainty of the measurement is $\Delta\delta = (\sqrt{N}T)^{-1}$, which is the standard quantum limit. The average signal of both even- and odd-$N$ cases (while still using the protocol optimized for odd $N$) is $\langle S_x \rangle = (\langle S_x \rangle_{even} + \langle S_x \rangle_{odd})/2$, corresponding to an average phase magnification of $N/2$, and the average variance of $S_x$ is $\Delta S_x^2 = [(\Delta S_x)_{even}^2 + (\Delta S_x)_{odd}^2]/2$. The uncertainty of the measurement including the same number of the even- and odd-$N$ trials is $\Delta\delta = [(N^2 + N)T^2/2]^{-1/2}$, which is approximately $\sqrt{2}(NT)^{-1}$ for $N \gg 1$ [53]. With an average phase magnification of $N/2$, the SCSP is still very robust against excess noise even if both the odd- and even-$N$ cases are considered. We get essentially the same result if we consider a protocol that is optimized for even $N$.

Although the SCSP has the advantage of high robustness, the Schrödinger cat (SC) state is very fragile against collisions with background atoms, thus potentially requiring the use of a cryogenically cooled vacuum chamber. As such, it is likely that the initial realization of a spin-squeezed CPT clock will make use of the SCSP with a value of the squeezing parameter that is far smaller than the value needed for realizing the SC state because, as we have shown in Ref. [31], the fragility of the spin-squeezed system (characterized by the reduction in the fringe contrast in the final signal as a result of collisions with background atoms) increases monotonically with increasing value of $\mu$ (up to $\mu = \pi/2$). In Ref. [31], we show that for the SCSP with a value of $\mu$ much smaller than $\pi/2$ (as small as $\sim 4\sqrt{2/N}$), the sensitivity can still approach the Heisenberg limit, within a factor of $\sqrt{2}$, for either parity of $N$. We recall from the preceding discussion that, for the case of the SCSP with $\mu = \pi/2$, the enhancement in sensitivity achievable is also smaller than the Heisenberg limit by a factor of $\sqrt{2}$ when averaged over odd and even values of $N$. Thus, in fact, the SCSP achieves the same sensitivity (namely the Heisenberg limit within a factor of $\sqrt{2}$) for a broad range of values of $\mu$: from $\sim 4\sqrt{2/N}$ to $\pi/2$.

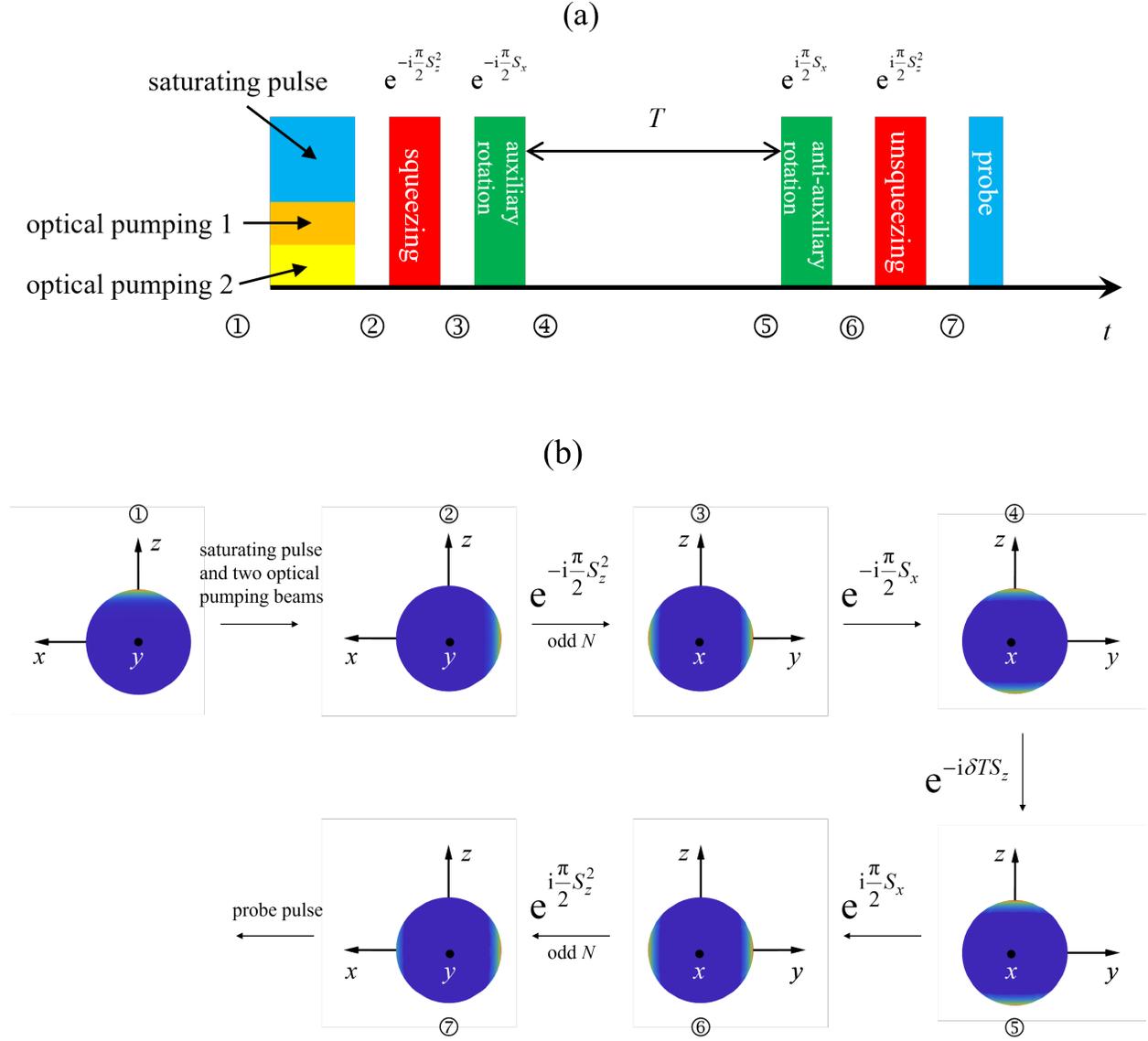

Figure 3: Schrödinger cat CPT clock (SC-CPT clock) protocol. (a) Pulse sequence of the SC-CPT clock. The first block is the combination of the saturating pulse and the two optical pumping beams. The second block is the squeezing pulse; the third block is the auxiliary pulse; the fourth block is the anti-auxiliary pulse; the fifth block is the unsqueezing pulse. The last block is the probe pulse. The states before and after each pulse are labeled from ① to ⑦. The Husimi quasi-probability distributions of the states ① to ⑦ are plotted in (b). (b1) All the atoms are in the $|\uparrow\rangle$ state initially. (b2) All atoms pumped into the CPT dark state after the saturating pulse. (b3) $y$-directed Schrödinger cat state generated by one-axis-twist squeezing. (b4) $z$-directed Schrödinger cat state after the auxiliary rotation. (b5) $z$-directed Schrödinger cat state with a phase shift after the dark period. (b6) $y$-directed the Schrödinger cat state with the phase shift after the anti-auxiliary rotation. (b7) State after the unsqueezing pulse.

The key difference between the SCSP with a small value of $\mu$ and the SCSP with $\mu = \pi/2$ is in the degree of robustness against excess noise. As shown in detail in Ref. [31], the degree of robustness for the SCSP protocol gradually increases with increasing values of $\mu$. We discuss next the possibility of using the echo squeezing protocol (ESP), which can also approach the Heisenberg limit of sensitivity, within a factor of $\sqrt{e}$, for a very small value of the squeezing parameter. We will then discuss the comparative advantages and disadvantages of the SCSP with a small value of $\mu$ and the ESP, regarding the flexibility of operating parameters and robustness against excess noise.

## 5. Echo-squeezed CPT clock

The so-called echo squeezing protocol (ESP) [30, 54] can also enhance the sensitivity of the CPT clock to approach the Heisenberg limit. Briefly, the steps of the ESP are similar to that of the Schrödinger cat sate protocol (SCSP). However, the few differences in steps between the ESP and the SCSP result in significant distinctions in behavior between these two protocols. Those distinctions will be summarized at the end of this section. The protocol for the echo-squeezed CPT clock (ES-CPT clock) is illustrated in Figure 4. We have shown the steps in Figure 4(a), and the corresponding Husimi QPD in Figure 4(b). As can be seen, the steps before the detection are the same as those for the SC-CPT clock (for odd values of $N$) except that in the ES-CPT protocol the optimal value of the squeezing parameter $\mu$ is $1/\sqrt{N}$, which, for a typically expected value of $N$ of experimental interest, is very small. At the detection stage of the ES-CPT protocol, we detect the population of atoms in the state $\left|\frac{\pi}{2},\frac{\pi}{2}\right\rangle$ with the Raman beams, which is equivalent to measuring the operator $S_y$. This is different from the SC-CPT protocol, for which we measure the state $\left|\frac{\pi}{2},0\right\rangle$, which is equivalent to measuring the operator $S_x$. In practice, this means that for the ES-CPT protocol, the phase of the clock VCO at the detection stage is ninety degrees shifted from that of the saturating pulse, as indicated in the last block of Figure 4(a). While the two protocols are similar in appearance, the quantum states during the intermediate steps are drastically different when compared to the SC-CPT with $\mu = \pi/2$, as can be seen from the Husimi QPDs illustrated in Figure 4 (b). However, as we have shown in Ref. [31] in a more generalized context, the intermediate states for the SC-CPT with a very small value of $\mu$ would be quite similar to those for the ES-CPT.

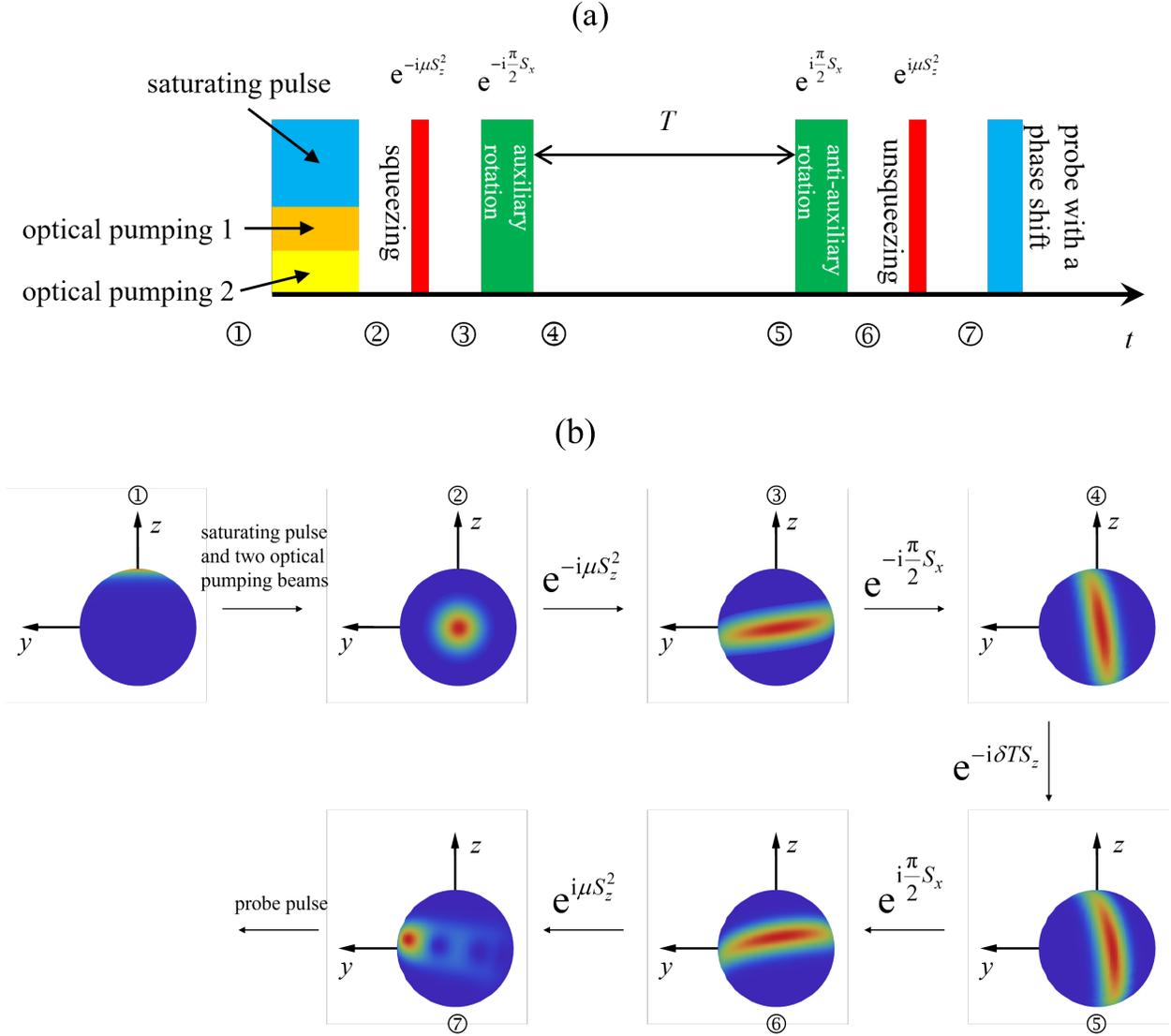

Figure 4: Echo squeezing CPT clock (ES-CPT clock) protocol. (a) The pulse squence for the ES-CPT protocol, which is the same as that for the SC-CPT clock in structure, but differ in some ways. A key differences between the ES-CPT and the SC-CPT are the values of $\mu$ for the squeezing and unsqueezing pulses. Furthermore, the phase of the probe pulse for the ES-CPT is different from that used for the SC-CPT, corresponding to detection of different quantum states. The optimal value of the echo squeezing protocol is $\mu = \mathrm{arccot}\sqrt{N-2}$, which is approximately $1/\sqrt{N}$ in the limit $N \gg 1$. (b) The corresponding Husimi QPDs, which are drastically different from those for the SC-CPT.

The sensitivity enhancement achievable by the echo squeezing protocol has also been investigated in detail [30]. As mentioned above, the optimal value of the squeezing protocol is $\mu = \text{arccot}\sqrt{N-2}$, which is approximately $1/\sqrt{N}$ in the limit $N \gg 1$. The uncertainty of the measurement with the optimal value of $\mu$ is $\sqrt{N(N-1)}\sin\mu\cos^{N-2}\mu$, which is approximately $\Delta\delta = \sqrt{e}(NT)^{-1}$ for $N \gg 1$; this represents a sensitivity of the Heisenberg limit within a factor of $\sqrt{e}$.

The signal of the ES-CPT clock $\langle S_y \rangle$ with respect to the phase shift $\delta T$ is plotted in Figure 5. Here, the solid curve is the signal of the ES-CPT clock for $N = 41$, while the dashed curve shows the signal for the corresponding conventional CPT clock. The signal of the ES-CPT has a periodicity of π, in contrast to the periodicity of 2π for the conventional CPT clock.

To discuss the enhancement in sensitivity under this protocol, and compare it with that for the conventional CPT clock in a transparent manner, it is useful to recall first the process used for determining the frequency shift in a conventional clock. As can be seen from the dotted trace in Figure 5, the signal for the CPT clock has a minimum when the phase-shift, or, equivalently, the detuning of the clock away from atomic resonance, is zero. If there is a shift in the central frequency of the VCO away from its ideal value that is resonant with the clock transition, the position of the signal minimum will shift. To determine the value of this shift, it is customary and optimal to use what can be called the hopping technique (which amounts to a square-wave modulation), as discussed in detail in Refs. [31] and [41]. Specifically, in this technique, the signal is measured at two frequencies (as indicated by the two vertical dotted lines in Figure 5) that are shifted in opposite directions away from the stationary value of the clock VCO frequency. The amplitudes of these frequencies are chosen such that the corresponding values of the phase shift (i.e., $\delta T$) are $\pm \pi/2$. The difference between these two measurements is considered the effective signal. Of course, if the stationary value of the clock VCO frequency is exactly on resonance, this signal will be null. However, if the stationary value of the clock VCO frequency shifts due to some unwarranted perturbation, the signal will be positive or negative, depending on the sign of the shift, thereby making it possible to determine the magnitude and sign of the shift. In practice, this signal is fed back to the clock VCO to keep it null-valued during the operation of the clock, thereby locking the stationary value of the clock VCO frequency to the atomic resonance.

Thus, the operator that is measured under the hopping technique for the conventional CPT clock can be expressed as $S_c = \left[ S_x(\delta + \pi/2T) - S_x(\delta - \pi/2T) \right]/2$, where we have inserted the division by two in order to enable the proper comparison with the ES-CPT protocol. Then the

signal can be written as $\langle S_c \rangle = (N/2)\sin\delta T$, and the noise can be expressed as

$\Delta S_c = \sqrt{\left[\left(\Delta S_x(\delta+\pi/2)\right)^2 + \left(\Delta S_x(\delta-\pi/2)\right)^2\right]/2} = \left(\sqrt{N}/2\right)|\sin\delta T|$. For convenience, we define a phase magnification factor (PMF) in a manner so that for the conventional CPT clock it has the value of unity at $\delta = 0$. It is easy to see that this is satisfied if the PMF is defined as $\partial_{\delta T}\langle S_c \rangle/(N/2)$; here, we have used the short-hand notation that $\partial_\theta f \equiv \partial f/\partial\theta$. From the expression above, it is easy to see that the corresponding noise at $\delta = 0$ is $\sqrt{N}/2$.

Consider next the protocol for the ES-CPT clock. In this case, we note that the signal is already asymmetric around $\delta = 0$. As such, it is not necessary to apply the hopping technique in this case. Instead, the signal of the ES-CPT clock is simply $\langle S_y \rangle$, and the noise is $\Delta S_y$. At $\delta = 0$, the phase magnification factor (PMF) is thus $\partial_{\delta T}\langle S_y \rangle/(N/2) = (N-1)\sin\mu\cos^{N-2}\mu$ [30], and the noise is $\sqrt{N}/2$ [30]. This PMF reaches its maximum value at $\mu = \text{arccot}\sqrt{N-2}$. For $N \gg 1$, the optimal value of $\mu$ is approximately $1/\sqrt{N}$ and the maximum PMF is approximately $\sqrt{N/e}$. Therefore, compared to the conventional CPT clock, the enhancement in sensitivity for the ES-CPT clock originates solely from phase magnification, by a factor of $\sqrt{N/e}$, with the noise remaining unchanged. This aspect of the echo squeezing protocol (considered for a conventional microwave clock) was noted earlier in Ref. [30] and [54]. However, in these references, the actual shape of the fringes for the echo squeezing protocol was not shown.

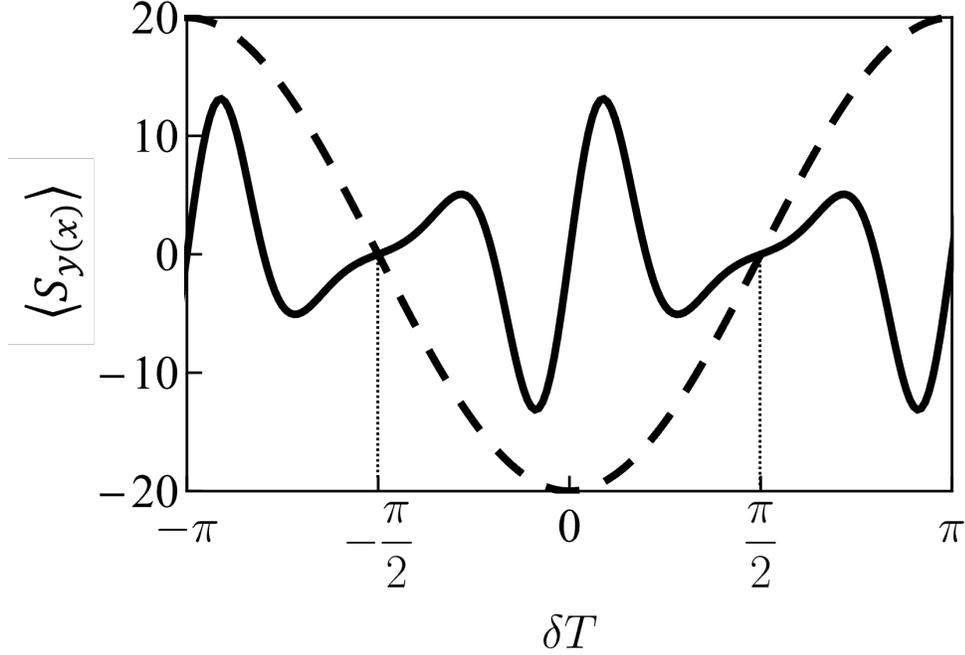

Figure 5. Signal of the ES-CPT clock (solid) and the conventional CPT clock (dashed) for $N = 41$. For the latter, it is customary and optimal to use what can be called the hopping technique, under which the signal is measured at two frequencies (vertical dotted line) that are shifted in opposite directions away from the stationary value of the clock VCO frequency. The amplitudes of these frequencies are chosen such that the corresponding values of the phase shift (i.e., $\delta T$) are $\pm \pi/2$. For the former, the hopping technique is not necessary since the signal is already asymmetric around the center.

We can now compare the behavior of these two protocols, and identify the differences between them. First, unlike the SCSP, the ESP produces the same result for *either* parity of $N$. Second, the maximum ideal enhancement in sensitivity under the ESP is lower than the Heisenberg limit by a factor of $\sqrt{e}$, and this occurs for the optimal value of $\mu = 1/\sqrt{N}$ for $N \gg 1$, while for the SCSP the maximum ideal enhancement in sensitivity, when averaged over both even-$N$ and odd-$N$ trials, is lower than the Heisenberg limit by a factor of $\sqrt{2}$, for a broad range of values of $\mu$, with the lowest value being $\sim 4\sqrt{2/N}$. We note that the averaging over the two parities of $N$ is only required for the SCSP for values of $\mu$ very close to $\pi/2$ [namely, for

$\left(\pi/2-\sqrt{2/N}\right)\leq\mu\leq\pi/2$]; for other values of $\mu$ that also yield the same sensitivity $\pi/2$ [namely, for $4\sqrt{2/N}\leq\mu\leq\left(\pi/2-\sqrt{2/N}\right)$], the signals are insensitive to the parity of $N$, thus requiring no averaging. Third, the sensitivities of these two protocols in the presence of potential excess noise including the one generated by the cavity-based OATS process are drastically different. This is because the optimal ESP magnifies the phase shift by a factor of $\sim\sqrt{N/e}$ while the SCSP with $\mu=\pi/2$ magnifies it by a factor of $N$. As a result, the net enhancement in sensitivity achievable for the SCSP with $\mu=\pi/2$, in the presence of the noise due to the cavity-based OATS process as well as any other potential excess noise, is much greater than the same for the ESP [29]. However, since the SCSP with $\mu=\pi/2$ requires a much larger value of the squeezing parameter $\mu$, and the Schrödinger cat state is very fragile against collisions with background atoms, it might be of greater practical interest to use the generalization of the SCSP [31], which works for a broad range of values of $\mu$ [namely, $4\sqrt{2/N}\leq\mu\leq\left(\pi/2-\sqrt{2/N}\right)$]. As we have shown in Ref. [31], while any value of $\mu$ in this range yields the same ideal enhancement in sensitivity, independent of the parity of $N$, the PMF increases monotonically with increasing $\mu$ (compensated by similarly increasing noise magnification). As such, the actual enhancement in sensitivity that can be achieved with the generalized SCSP, when taking into account the presence of the noise due to the cavity-based OATS process as well as any other potential excess noise, is expected to increase monotonically with $\mu$ (up to $\mu=\pi/2$). On the other hand, as we have shown in Ref. [31], the fragility of the spin-squeezed system (characterized by the reduction in the fringe contrast in the final signal as a result of collisions with background atoms) increases monotonically with increasing value of $\mu$ (up to $\mu = \pi/2$). Thus, the optimal choice of $\mu$ to be used, for the generalized SCSP, would depend on the expected rate of collisional loss.

The sensitivity in the presence of excess noise can be expressed as $\text{PMF}/\sqrt{\Delta S_{\text{QPN}}^2 + \Delta S_{\text{ES}}^2}$. The lowest excess noise reported for the cold-atom-based CPT clock of Ref. [14], which does not employ spin squeezing, is $50/\sqrt{N}$ with $N = 5 \times 10^6$. The corresponding sensitivity [i.e., $(\Delta \delta T)^{-1}$] is then $45$. For the ESP, the denominator of the sensitivity expression does not change while the numerator is increased by a factor of $\sqrt{N/e} \approx 1360$, which represents the net enhancement in sensitivity over what was observed in this experiment, under ideal conditions. The net sensitivity is enhanced to a value of $45 \times 1356 = 6.1 \times 10^4$, which is ~27.3 times the standard quantum limit. For the SCSP with $\mu = \pi/2$, we have $\text{PMF} = N/2$ and $\Delta S_{\text{QPN}} = \sqrt{N/2}$. Accordingly, the net sensitivity is enhanced to $0.9995 \times N/\sqrt{2}$, very close to the maximum achievable sensitivity. The net enhancement in sensitivity over what was observed in the experiment would thus be $\sim 7.85 \times 10^4$, if the squeezing process is ideal. Of course, the excess noise introduced by the squeezing process may limit the maximum enhancement achievable, as discussed further in the next section.

## 6. Conclusion and discussion

The technique of spin squeezing can be applied to the Ramsey CPT clock but cannot be used for the continuous light CPT clock. Nevertheless, the adaptation necessary for the CPT clock is not obvious because the Ramsey CPT clock is not trivially equivalent to the conventional Ramsey microwave clock. It is well known that there is an important difference between the dark state produced at the end of the saturating CPT pulse (which is the first of the two CPT pulses) and the state produced with a microwave π/2 pulse. The concrete protocols we describe are the Schrödinger cat CPT (SC-CPT) scheme, the generalization thereof for a broad range of values of the squeezing parameter, $\mu$, and the echo squeezing CPT (ES-CPT) scheme. The SC-CPT clock and the generalization thereof can reduce the uncertainty of the measurement of the frequency by a factor of $\sqrt{N/2}$, and the ES-CPT clock by a factor of $\sqrt{N/e}$. Thus, they both can get close to the Heisenberg limit in the absence of excess noise. However, in the presence of excess noise, the SC-CPT clock for $\mu = \pi/2$ can achieve a much higher sensitivity than the ESP does because it magnifies the phase shift to a much greater extent. Furthermore, the generalized form of the SC-CPT clock can achieve the Heisenberg limit of sensitivity, within a factor of $\sqrt{2}$, for a wide range of values of the squeezing parameter, ranging from $\mu = 4\sqrt{2/N}$ to $\mu = \pi/2$. We have also shown that the robustness of the generalized SC-CPT clock against excess noise, including those from the squeezing process itself, increases monotonically with the squeezing parameter. At the same time, the fragility of the squeezed quantum state, against collisions with background atoms, also increases monotonically with the squeezing parameter. As such, the optimal value of the squeezing parameter to be used for the SC-CPT clock would depend on the expected degree of collisional loss in the experiment.

The effect of the noise induced by the cavity-based one-axis-twist squeezing process, in the case of the Schrödinger cat state protocol, can be found in Ref. [29], and would be applicable for the same protocol applied to the CPT clock. Specifically, for a single-atom cooperativity of 0.9, corresponding to the particular set of operating parameters mentioned in Section 4, and one million atoms, the enhancement of sensitivity, as seen from the lower left graph of Fig. 6 in Ref.

[29] is ~28.4 dB, which is only 1.6 dB below the Heisenberg limit of 30 dB [55]. The corresponding noise analysis in the case of the echo squeezing protocol is shown in Ref. [c]. For the same parameters, the enhancement of sensitivity, as seen from the lower-left graph of Fig. 4 in Ref. [29], is ~12.5 dB, which is ~17.5 dB below the Heisenberg limit of 30 dB. As discussed in detail in Ref. [29], the Schrödinger cat state protocol is more resistant to any excess noise, including those resulting from non-idealities of the cavity-induced squeezing process. This is due to the fact that the quantum projection noise in this case is amplified by a factor of $\sqrt{N}$, thus dwarfing the effect of other noise. Of course, this increase in the quantum noise is counterbalanced by the fact that the phase is magnified by a factor of $N$, with the net enhancement in sensitivity being $\sqrt{N}$. For the generalized form of the SC-CPT clock with a value of the squeezing parameter ranging from $\mu = 4\sqrt{2/N}$ to $\mu = \pi/2$, the suppression of the effect of the excess noise from the squeezing process would increase monotonically with the value of $\mu$. The actual degree of enhancement achievable for $\mu < \pi/2$ requires extensive numerical modeling, which will be carried out in the near future.

Compared to the microwave clock, the CPT clock may experience additional sources of noise. The most important source of such additional noise is the light shift [26,40]. The absolute light shift can be expressed as $\Omega^2/2\Delta$, where $\Omega$ is the one-photon Rabi frequency and $\Delta$ is the common detuning. Thus, fluctuations in the laser intensity and the frequency can cause errors in the measured clock frequency. In Refs. [26,40], it has been shown that the light shift can be virtually eliminated if the populations of the two ground states before the initial CPT pulse are equal. Furthermore, a high degree of intensity and frequency stabilizations of the lasers can also suppress the light shift, to an arbitrarily small value in principle. All these techniques for suppressing the light shift are fully compatible with the squeezing protocols presented here.

## Acknowledgement:

This work has been supported equally in parts by the Department of Defense Center of Excellence in Advanced Quantum Sensing under Army Research Office grant number W911NF202076, and the U.S. Department of Energy, Office of Science, National Quantum Information Science Research Centers, Superconducting Quantum Materials and Systems Center (SQMS) under contract number DE-AC02-07CH11359.